\documentclass[aip,jcp,reprint,amsmath]{revtex4-1}
\usepackage{graphicx,color,soul}
\usepackage{xcolor}

\newcommand{\ta}{\tau_\alpha}
\newcommand{\tobs}{t_\text{obs}}
\newcommand{\mean}[1]{\langle #1\rangle}
\newcommand{\TK}{T_\text{K}}
\newcommand{\Tg}{T_\text{g}}
\newcommand{\To}{T_\text{on}}
\newcommand{\Tc}{T_\text{c}}
\newcommand{\Ts}{T_\text{s}}

\begin{document}

\title{Dynamical Phase Transitions and their Relation to Structural and Thermodynamic Aspects of Glass Physics}

\author{C. Patrick Royall}
\email{chcpr@bristol.ac.uk}
\affiliation{H.H. Wills Physics Laboratory, Tyndall Avenue, Bristol, BS8 1TL, UK}
\affiliation{School of Chemistry, University of Bristol, Cantock's Close, Bristol, BS8 1TS, UK}
\affiliation{Centre for Nanoscience and Quantum Information, Tyndall Avenue, Bristol, BS8 1FD, UK}

\author{Francesco Turci}
\affiliation{H.H. Wills Physics Laboratory, Tyndall Avenue, Bristol, BS8 1TL, UK}
\affiliation{Centre for Nanoscience and Quantum Information, Tyndall Avenue, Bristol, BS8 1FD, UK}

\author{Thomas Speck}
\affiliation{Institut f\"{u}r Physik, Johannes Gutenberg-Universit\"{a}t Mainz, Staudingerweg 7-9, 55128 Mainz, Germany}

\begin{abstract}
We review recent developments in structural-dynamical phase transitions in trajectory space based on dynamic facilitation theory. An open question is how the dynamic facilitation perspective on the glass transition may be reconciled with thermodynamic theories that posit collective reorganization accompanied by a growing static length scale, and eventually, a vanishing configurational entropy. In contrast, dynamic facilitation theory invokes a dynamical phase transition between an \emph{active phase} (close to the normal liquid) and an \emph{inactive phase} which is glassy, and whose order parameter is either a \emph{time-averaged} dynamic or structural quantity. In particular, the dynamical phase transition in systems with non-trivial thermodynamics manifests signatures of a lower critical point that lies between the mode-coupling crossover and the putative Kauzmann temperature, at which a thermodynamic phase transition to an ideal glass state would occur. We review these findings and discuss such criticality in the context of the low-temperature decrease of configurational entropy predicted by thermodynamic theories of the glass transition. 
\end{abstract}

\maketitle


\section{Introduction}

Understanding the physical origin of the glass transition is a longstanding challenge in condensed matter physics. Cool any liquid sufficiently fast and it will not order into a crystal but remain a liquid before, eventually, it falls out of equilibrium and becomes a \emph{glass}. The single most important quantity when talking about supercooled liquids is the structural relaxation time $\ta$, which measures the \emph{average} time over which atoms and molecules rearrange such as to lose memory of their initial positions. Over time, a variety of theoretical approaches have been developed to account for the massive slowdown ($\ta$ increases by 14 orders of magnitude) in the dynamics of such supercooled liquids as they approach the experimental glass transition temperature $\Tg$~\cite{debenedetti,dyre2006,cavagna2009,berthier2011,berthier,biroli2013,ediger2012,berthier2016pt,royall2018jpcm}.

Among the approaches to explain this challenge is \emph{dynamic facilitation} \cite{speck2019,chandler2010}. In this Perspective, we offer a viewpoint that, while some aspects of dynamic facilitation may seem at odds with theories which posit a thermodynamic origin to the glass transition, in fact in a number of atomistic models, key predictions of facilitation may be reconciled with a thermodynamic interpretation. That is to say, \emph{the problem that we seek to address is how to reconcile dynamic facilitation with other, thermodynamic, theories of the glass transition.} We do not attempt to \emph{justify} dynamic facilitation, rather to try to link it with certain observations to thermodynamic interpretations of the glass transition. We refer the reader to refs.~\citenum{speck2019,chandler2010} and ~\citenum{biroli2013} for reviews of dynamic facilitation.

We begin by noting a few salient points from various theoretical treatments pertinent to our discussion (Sec.~\ref{sectionThermo}) and then discuss dynamical phase transitions between an \emph{active} phase (close to the normal liquid) and an \emph{inactive} phase in the context of dynamic facilitation (Sec.~\ref{sectionDynfac}). In Sec.~\ref{sectionSMu}, we move on to consider the so-called $\mu$-ensemble, which uses a time-averaged structural quantity to drive a dynamical phase transition \cite{speck2012}. We review the reweighting of configurations generated in the $\mu$-ensemble to access states representative of those very deep in the energy landscape~\cite{turci2017prx} (Sec.~\ref{sectionReweighting}). In Sec.~\ref{sectionAllTogether}, we discuss the interpretation of these results within the picture of a supercooled liquid and crystal branch in a plot of (configurational) entropy and temperature originally introduced by Kauzmann \cite{kauzmann1948}, as sketched in Fig. \ref{figCavagnaWDIAM}.

Put briefly, we conclude that inactive phases have lower configurational entropy than the normal liquid, and that at sufficiently low temperature the inactive phase should merge with the active phase. The temperature at which this occurs may be reasonably close to (but is expected to be above) the Kauzmann temperature. More specifically, in the systems we consider~\cite{speck2012,turci2017prx,pinchaipat2017,turci2018epje,campo2020}, this merging of the active and inactive phase appears to lie between the Kauzmann temperature and the mode-coupling crossover. We discuss numerical evidence that this merging is controlled by a lower critical point. One consequence of the perspective we offer here is that the dynamical phase transition invoked in dynamic facilitation theory can, for suitable systems, be interpreted within the context of the ``Kauzmann plot'' of configurational entropy as a function of temperature (Fig. \ref{figCavagnaWDIAM}). Therefore, despite the very different starting points we offer a means to reconcile approaches of dynamic facilitation and those based on a thermodynamic interpretation. Before concluding, in Sec.~\ref{sectionConclusions}, we consider the challenges of this approach and provide an outlook for future work in Sec. ~\ref{sectionChallengesOutlook}.

\section{Key Points from Thermodynamic Theories}
\label{sectionThermo}

Compared to crystallisation, little apparent change in two-point structure assumed by the constituent particles occurs during vitrification~\cite{royall2015physrep}, suggesting that while the origin of the dynamic slowdown could be due to a phase transition related to some kind of change in amorphous order or structure~\cite{bouchaud2004}, the case that the glass transition is a predominantly dynamical phenomenon is compelling \cite{chandler2010}.

In the limit of infinite-dimensional systems, structural correlations beyond two-point correlations become irrelevant. In this limit mean-field treatments become exact, see ref.~\citenum{charbonneau2017} for a more complete description. Upon cooling (or, in the case of hard-sphere-like systems, compression) a dynamical transition occurs, related to the mode-coupling transition~\cite{goetze1999,charbonneau2005}. At deeper supercooling (compressing further for hard-sphere-like systems), a thermodynamic transition to a state with sub-extensive entropy, a so-called \emph{ideal glass}, is encountered. This thermodynamic transition, along with the dynamic transition are both captured by mean-field random first order transition theory (RFOT)~\cite{lubchenko2007}.

While the situation in high dimension is now understood, back in dimension $d=3$, a full explanation remains elusive. What we do know is that the early stages of glassy dynamics (i.e. the first few decades of increase in relaxation time with supercooling) are accurately described by mode-coupling theory (MCT)~\cite{goetze,goetze1999,charbonneau2005}. As input, MCT calculations utilize structural two-point correlations and provide evolution equations for dynamic two-point correlations. However, unlike the case in high dimension, in lower dimensions the MCT approach fails when many-body correlations become important for relaxation, which occurs typically after 4-5 decades of increase in structural relaxation time~\cite{brambilla2009,hallett2018}. Recently, progress has been made in this direction \cite{biroli2006,szamel2013}, and a generalised MCT approach has provided a route to address this limitation, although the Herculean task of solving the higher-order coupled equations should not be understated~\cite{janssen2015}.

At deeper supercooling in $d=3$, past the (avoided) mode-coupling transition, random first order transition theory (RFOT)~\cite{lubchenko2007} and Adam-Gibbs theory~\cite{adam1965} both feature a vanishing configurational entropy at non-zero temperature as is found in high dimension~\cite{charbonneau2017}. This implies a corresponding divergent \emph{static} correlation length~\cite{montanari2006}. The qualitative picture is that of cooperative motion of more and more particles in order for the liquid to relax.

Approaches based on replica theory imagine multiple coupled copies or ``replicas'' of the same system. A field $\varepsilon$ is applied, which favours the \emph{overlap} of particle positions between different replicas. At high temperature, in the liquid, there is zero overlap between different replicas, but upon cooling a phase transition occurs to states which feature high overlap. In other words, it becomes favourable for states with similar configurations to be found. This may be taken as a further manifestation of the drop in configurational entropy~\cite{parisi2010}.

This application of an external field to induce a phase transition \cite{berthier2013,turner2015} has some parallels with the the $s$- and $\mu$-ensembles that we discuss in the following \cite{chandler2010,hedges2009,speck2012}. However the application of an external field $\varepsilon$ in the context of replica theory is fundamentally different from the external field--induced transitions that we discuss here, because these occur in \emph{trajectory space} whereas replica theory concerns \emph{configurations}, not \emph{trajectories}.

Other theories, such as geometric frustration, emphasize \emph{locally favoured} -- or locally \emph{preferred -- structures} (LFS) \cite{tarjus2005}. These locally favoured structures are geometric motifs that are minima of the local (free) energy. Their concentration appears to increase as a glass-former is cooled down, and they have been identified with the emergence of slow dynamics~\cite{royall2015physrep,coslovich2007,royall2008,leocmach2012} and a drop in configurational entropy \cite{hallett2018,hallett2020}. Related to these observations, the geometric frustration theory imagines an avoided phase transition to a state of LFS which would occur e.g. in curved space~\cite{tarjus2005}. The increase in LFS with supercooling, while suppressed in Euclidean space~\cite{tarjus2005,turci2017prl} is nevertheless understood to be compatible with the picture of some thermodynamic transition where configurational entropy becomes small~\cite{tarjus2005}.

\begin{figure}[t]
\centering \includegraphics[width=85mm]{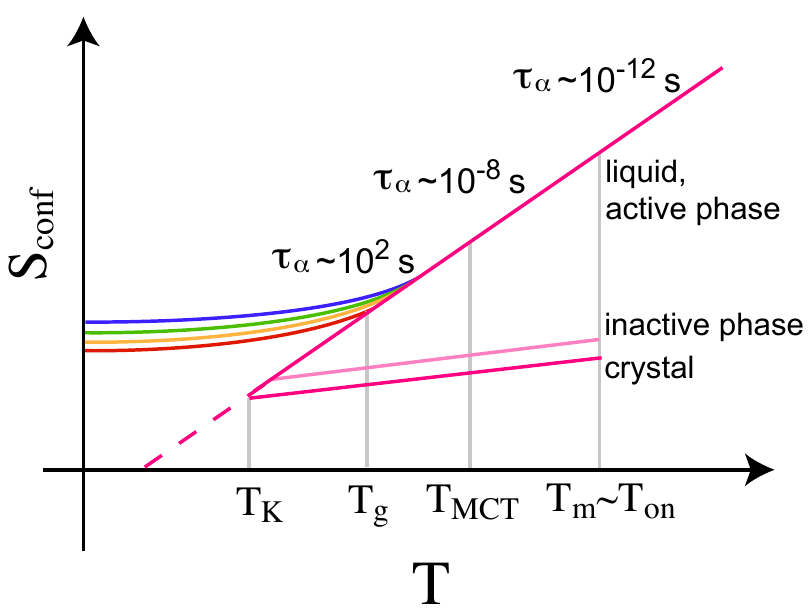} 
\caption{Roadmap to the glass transition for structural glasses. Configurational entropy $S_\mathrm{conf}$ as a function of temperature. Typically, configurational entropy of liquids falls faster than that of crystals as a function of temperature. This suggests that, at some low temperature  -- the Kauzmann Temperature $\TK$ -- the liquid configurational entropy would fall below that of the crystal. Note that we imagine some population of defects in the crystal, so that its configurational entropy is itself non-zero ~\cite{stillinger2001,royall2018jpcm}. $\Tg$ is the operational glass transition temperature where the structural relaxation time reaches 100 s. $T_\text{MCT}$ is the mode-coupling transition. $T_\text{m}$ is the melting point and $\To$ denotes a crossover temperature below which relaxation occurs through local fluctuations. The two branches of the dynamical phase transition of the $\mu$-ensemble (see section \ref{sectionSMu}) are indicated. The active phase is effectively indistinguishable from the equilibrium liquid. The inactive, or structure-rich phase has a lower configurational entropy, which \emph{we presume} lies close to, but is slightly larger than, that of the crystal. The putative lower critical point of the active and inactive phases is thus bounded by the liquid and crystal and thus lies at higher temperature than the Kauzmann point.}
\label{figCavagnaWDIAM} 
\end{figure}

The low-temperature fate of structural glassforming systems in low dimension is thus summarised in Fig. \ref{figCavagnaWDIAM}. We do \emph{not} enter into a discussion of the nature of any ideal glass transition at $\TK$ here, or indeed its possible avoidance, but refer the reader to refs.~\cite{royall2018jpcm,tanaka2003,stillinger2001,kauzmann1948}. Similarly, while the configurational entropy is a challenging quantity to define, let alone measure \cite{osawa2018}, we save this discussion for section \ref{sectionReweighting}.
We see that the configurational entropy of the liquid and crystal are expected to become equal at the Kauzmann temperature upon extrapolation~\cite{kauzmann1948}. Here we shall argue that the dynamical phase transition of facilitation leads to two dynamical phases whose configurational entropies are bounded by the crystal and liquid as shown in Fig. \ref{figCavagnaWDIAM}. To proceed, in the next section we discuss the dynamic facilitation approach.

\begin{figure*}[!htb]
\centering \includegraphics[width=120mm]{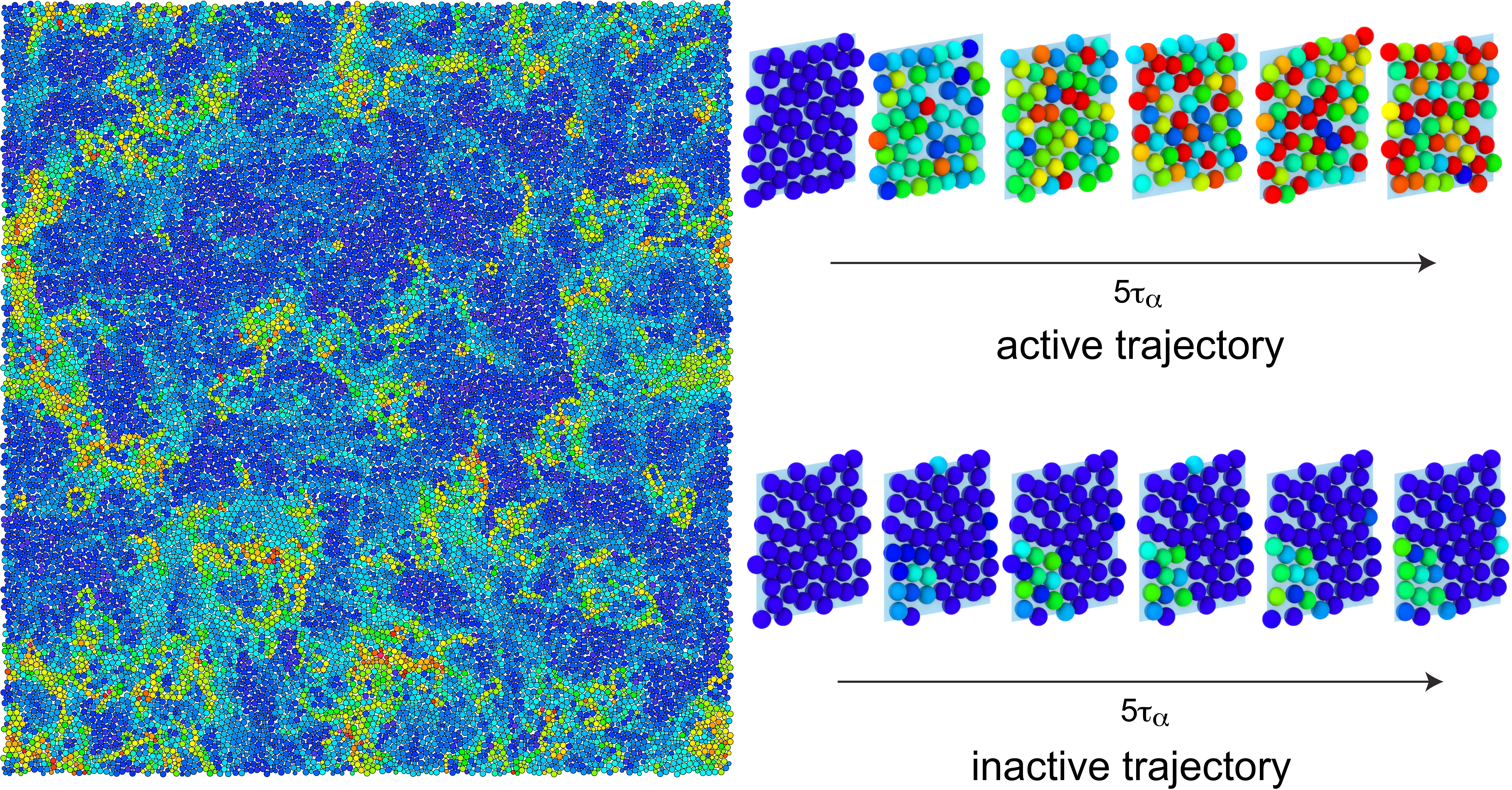} 
\caption{Dynamic heterogeneity and trajectories. Main panel: Dynamic heterogeneity in a simulation of binary hard discs with a size ratio of 1:1.4. Blue particles have moved the least (mean squared displacement $\langle r^2 \rangle <0.01\sigma_\mathrm{large}^2$) and red have moved the most ($\langle r^2 \rangle>\sigma_\mathrm{large}^2$). Here the timescale is taken over the structural relaxation time $\tau_\alpha$. Right is shown a schematic example of an active trajectory (top) and inactive trajetcory (bottom).}
\label{figTrajectories} 
\end{figure*}


\section{Dynamic Facilitation}
\label{sectionDynfac}

In dynamic facilitation theory~\cite{chandler2010}, dynamic arrest is attributed to emerging \emph{kinetic} constraints. This theoretical approach focuses on the role of real-space fluctuations and \emph{dynamic heterogeneities}~\cite{harrowell2011}, i.e., the coexistence of mobile and solid-like regions in a supercooled liquid, instead of thermodynamic and structural mechanisms. Like several competing theories, dynamic facilitation is built around the idea of a phase transition, but here its nature is profoundly different: in contrast to conventional thermodynamic phase transitions, where the coexisting phases are characterised by distinct static properties (e.g. the density difference in liquid-gas coexistence), dynamical facilitation starts from the observation that in the supercooled liquid one encounters regions that relax (warm-coloured regions in Fig.~\ref{figTrajectories}) on the timescale of the structural relaxation time $\ta$, and some that do not (blue regions in Fig.~\ref{figTrajectories}). Both of these \emph{coexist} with one another. The kinetic constraints govern the effective dynamics of localized excitations that sustain motion within the mobile regions.

\begin{figure}[!t]
\centering 
\includegraphics[width=60mm]{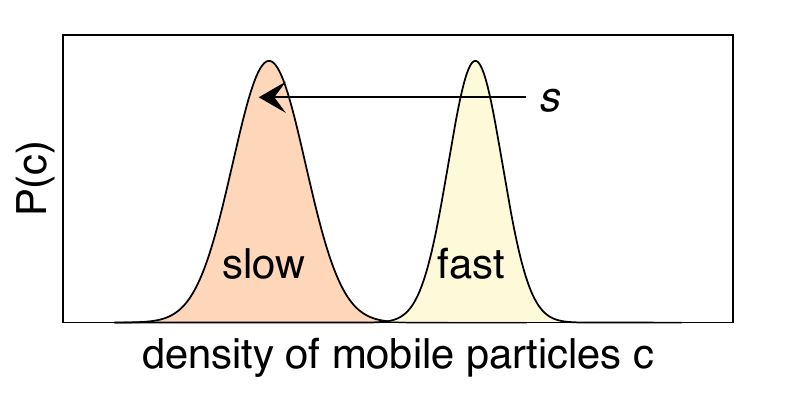} 
\caption{Schematic of the dynamical phase transition between \emph{inactive}  (slow) and \emph{active} (fast) \emph{trajectories}. Here $c$ is the density of mobile particles per trajectory, characterised by larger values of summed short-time particle displacements. The application of the field $s$ brings the active and inactive populations of trajectories to a dynamical phase coexistence.}
\label{figDynamicTransitionEasy} 
\end{figure}

To be more specific, let us decompose a large system into distinct subsystems containing a few hundred particles. Following the dynamics for an observation time $\tobs$ on the order of a few structural relaxation times, we find that the resulting \emph{trajectories} performed by some subsystems will display large cumulative displacements, while others will show very little change in their particle positions. Such behaviour can be interpreted as the manifestation of a \emph{dynamical phase transition} between a relatively fast-moving (\emph{active}) state of trajectories and a slow-moving (\emph{inactive}) state (cf. the trajectories in Fig.~\ref{figTrajectories}). In a suitable dynamical ensemble, it can be shown that a genuine \emph{dynamical phase coexistence} can be established, and that a bimodal distribution of active and inactive trajectories emerges~\cite{hedges2009,speck2012,campo2020}. Such a dynamical phase coexistence of slow/inactive and fast/active \emph{trajectories} is indicated in Fig. \ref{figDynamicTransitionEasy}. The analogy to conventional phase transitions is made by identifying the density of mobile particles $c$ with density in a liquid-gas transition, for example.

\subsection{Excitations: the elementary units of relaxation}

It is important to consider the elementary units of relaxation. The interpretation here is profoundly different to the cooperatively rearranging regions with an increasing lengthscale postulated by certain thermodynamic approaches. Facilitation places much emphasis on the mobility of individual particles (or small groups of particles). It notes that motion in deeply supercooled liquids occurs through local events termed \emph{excitations} whose timescales are much shorter than the (overall) relaxation time. Upon cooling, these relaxation events become rarer, but remain essentially unchanged. Thus, an increasing dynamic lengthscale $\xi_\mathrm{fac}$ corresponds to larger separations between these events as their population falls.

Coupling between excitations is achieved through ``surging'' events, which are long-ranged, string-like motions of very small displacements (around 0.1 particle diameters). The motion is very often reversed, such that many surging events are required before another excitation is ``facilitated''~\cite{gebremichael04}. Through surging events, excitations are coupled to one another logarithmically such that the activation energy for relaxation follows $E_\mathrm{fac} \sim \log \xi_\mathrm{fac}$~\cite{chandler2010}. Now the Boltzmann factor implies that the concentration of excitations falls like $c \sim \exp(- E_\mathrm{fac}/k_BT)$, so $\log c\propto 1/T$. The mean separation between excitations is $\xi_\mathrm{fac} \approx c^{-1/d}$ where $d$ is the spatial dimension. Thus, to leading order, the activation energy scales as $1/T$ and the timescale for relaxation $\log \tau_\alpha \sim E_\mathrm{fac}/k_BT\sim\exp(1/T^2)$. These arguments underlie the Elmatad-Garrahan-Chandler form for the relaxation time~\cite{elmatad2009}
\begin{equation}
  \log \tau_\alpha = \left( \frac{J}{\To}  \right)^2  \left(\frac{\To}{T}-1\right)^2
\label{eqElmatad}
\end{equation}
where $\To$ is the onset temperature for slow dynamics and $J$ is a parameter to scale the activation energy. A range of glassformers with varying chemical properties have been shown to collapse onto a single curve described by Eq. \ref{eqElmatad} which fits the data at least as well as the semi-empirical Vogel-Fulcher-Tamman form~\cite{elmatad2009}.

\begin{figure}
\centering
\includegraphics{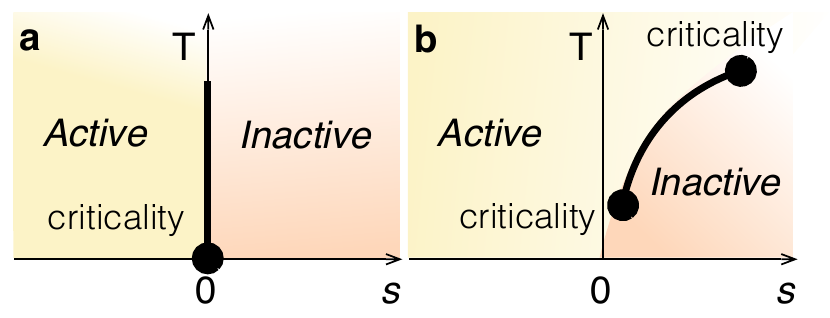}
\caption{Dynamical phase transitions in kinetically constrained models. (a) The East Model (KCM), which undergoes a dynamical transition at $s=0$ for all temperatures terminating in a critical point at $\Tc=0$. (b) Adding ``softness'' to the East Model shifts the transition and moves the lower critical point to a finite temperature $\Tc>0$ and positive field $s_\text{c}>0$. There is now also an upper critical point beyond which dynamics is no longer facilitated. Adapted from Ref.~\citenum{elmatad2013}.}
\label{figSoftEast}
\end{figure}

\vspace{10mm}
\subsection{Kinetically Constrained Models}

In a series of papers, idealised lattice models of supercooled liquids, so-called \emph{kinetically constrained models} (KCMs)~\cite{ritort2003,merolle2005,garrahan2007,charbonneau2008,elmatad2013} have been investigated. These models capture the essence of dynamical facilitation as they essentially neglect the details of particle-particle interactions, focusing on the hindrance to relaxation that is a distinctive feature of supercooled liquids. This is represented through simple and idealised on-lattice rules for the motion of particles or the relaxation of spin excitations. Although the Hamiltonian of such models is often designed to be trivial, they exhibit highly nontrivial, glassy dynamics, in particular faster than exponential (i.e. super-Arrhenius) increase of the relaxation time with decreasing temperature and dynamic heterogeneities. The fact that these idealized models exhibit so much of the phenomenology of dynamic arrest in liquids (such as super-Arrhenius relaxation and dynamical heterogeneity) provides strong evidence that a thermodynamic glass transition -- absent here by construction -- is not necessarily required.

Regardless of the details of their respective kinetic rules, kinetically constrained models show a dynamical first-order phase transition between an ``active'' phase (many spin relaxations) and an ``inactive'' phase, which is stuck in jammed configurations for long times. The phase diagram is sketched in Fig.~\ref{figSoftEast}(a), with a coexistence line that, in the thermodynamic limit, emanates from a critical point at $\Tc=0$ and lies at $s=0$. The dynamical field $s$ which drives the transition pertains to the time-averaged mobility along a trajectory and is discussed in detail in Sec \ref{sectionSMu}.

This picture holds for ``hard'' constraints that cannot be violated. In atomistic systems the corresponding effective kinetic constraints are emergent and possibly can be violated. This has been included in KCMs in the form of an energy barrier, and overcoming this barrier allows to bypass the constraints~\cite{elmatad2013,gutierrez19}. The phase diagram now changes as sketched in Fig.~\ref{figSoftEast}(b): While there is still a coexistence line delineating the active and inactive phases, it bends away from the temperature axis with $s_\text{coex}(T)>0$. The lower critical point moves to a finite temperature, and now there is an upper critical point terminating coexistence. Beyond the two critical points facilitation is weak: at low temperatures it becomes favorable to ``pay'' the energy cost since relaxation through the constrained dynamics is taking even longer, while at high temperatures the thermal energy overwhelms the constraints. In the following we present evidence that this qualitative picture carries over to dynamic phase transitions in atomistic model glass formers.

A more formal understanding of the \emph{$s$-ensemble} relies on the application of theory of large deviations to non-equilibrium steady states, where the transition is determined from the non-analyticities of cumulant generating functions that can be interpreted as the non-equilibrium analog of free energies~\cite{touchette2009}. Interestingly, such dynamical transitions seem to be robust to the application of external driving forces that break detailed balance~\cite{speck2011,turci2011,turci2012}.

\vspace{10mm}
\subsection{Simultaneous ``success'' of thermodynamic and dynamic approaches}

At this point it is helpful to recall that both thermodynamic approaches \emph{in low dimension} and dynamic facilitation are ``phenomenological theories'' that aim to capture the dominant mechanism through which relaxation in supercooled liquids is hampered. With the exception of mode-coupling theory, we are not dealing with first-principles microscopic theories that yield explicit expressions, which makes it hard to discriminate both approaches based on experimentally accessible data and computer simulations. That both thermodynamic and dynamic interpretations describe the available data thus leaves us with a conundrum: how can the observed --- albeit subtle --- structural changes occurring in glass forming liquids \cite{royall2015physrep,tanaka2019} (consistent with thermodynamic Adam-Gibbs or RFOT approaches) be compatible with the picture emerging from dynamical facilitation?

\begin{figure}[t]
\centering \includegraphics[width=0.8\columnwidth]{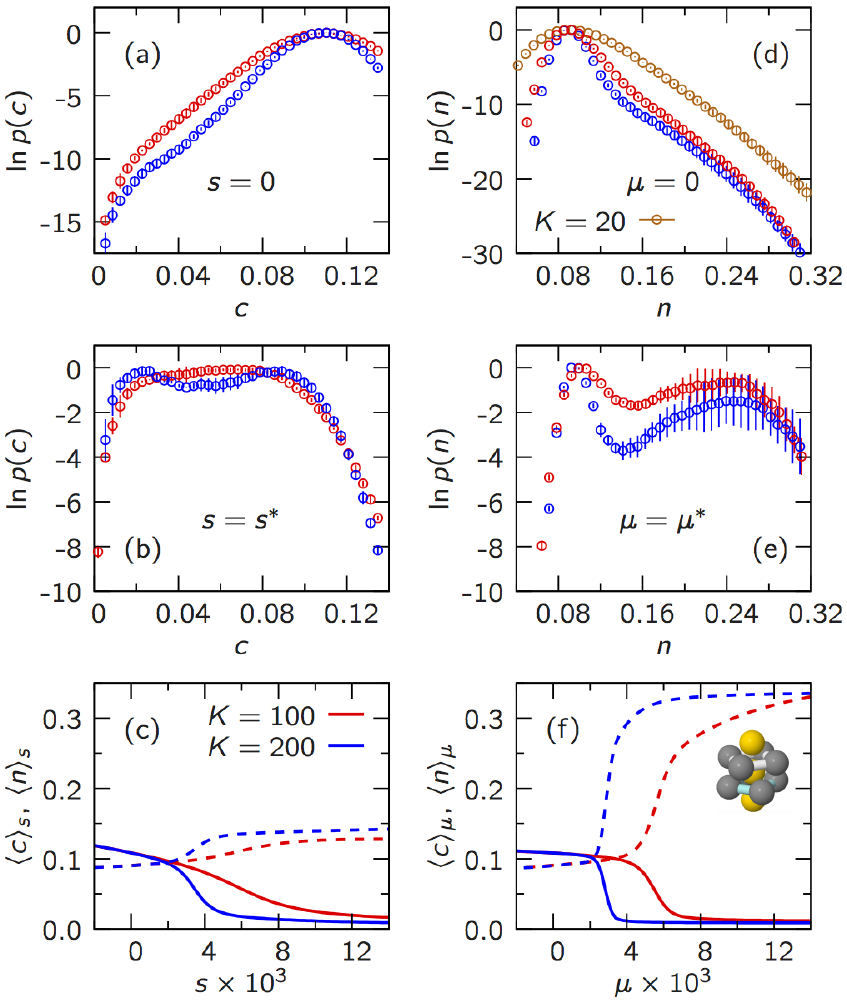} 
\caption{Phase transitions in trajectory space for the Kob-Andersen binary Lennard-Jones mixture. Throughout, red and blue lines refer to trajectory lengths $\tobs\approx5\ta$ and $\tobs\approx10\ta$, respectively. Left column: $s$-ensemble. (a)~Probability distributions $p(c)$ for the density of mobile particles $c$ for two trajectory lengths. The non-concave shape indicates a phase transition in trajectory space as becomes obvious from the bimodal distribution~(b) at the field $s^\ast$ that maximises the fluctuations $\langle{c^2}_s\rangle-\langle{c}_s\rangle^2$. (c)~Average fractions of mobile particles (solid lines) and bicapped square antiprism cluster population (dashed lines) \textit{vs.} the biasing field $s$. Right column: (d-f)~as left column but for the $\mu$-ensemble. Here the bicapped square antiprism (depicted in (f)) is the locally favoured structure. Reproduced with permission from Ref.~\citenum{speck2012}.}
\label{figSMu}
\end{figure}

Evidence in support of each approach has been presented. For example, several recent studies~\cite{royall2018jpcm} have shown that the deeply supercooled state presents a low configurational entropy that monotonically decreases with temperature as expected in the Adam-Gibbs/RFOT scenario, both in advanced Monte-Carlo simulations~\cite{berthier2017pnas} and colloidal experiments~\cite{hallett2018,gokhale2016,hallett2018,hallett2020}, as well as a growing static length scale in experiments on molecules~\cite{albert2016}. On the other side, support for dynamic facilitation comes from computer simulations of atomistic models (including three dimensional Lennard-Jones binary mixtures and hard spheres \cite{keys2011,isobe2016,thompsonthesis}) which are less idealized than KCMs and in which kinetic constraints are not present by construction but emerge from interparticle forces. While such simulations, of course, cannot conclusively prove the validity of dynamical facilitation, the absence of a dynamical phase transition would have been a considerable blow to the theory, suggesting that it might be limited to KCMs. Some success was found even in experiments with molecular systems, where KCMs were shown to explain calorimetric effects in the glass transition \cite{keys2013}, while colloids exhibit the dynamical phase transition of Fig. \ref{figSMu} \cite{pinchaipat2017,abou2018}.

Although originating from studies of kinetically constrained models, dynamical phase transitions are not an exclusive trait of dynamic facilitation. It appears that any sufficiently complex model with long-lived metastable states can be driven into an inactive phase using an order parameter that couples to mobility. In particular, spin glasses were shown to exhibit a dynamical active-inactive transition~\cite{jack2010}. This is significant as these models are amenable to mean-field theory and some follow the same physics as structural glasses in high-dimension \cite{charbonneau2017,berthier2019jcp}. The suggested phase diagram of one-step replica symmetry-breaking models is more complicated than Fig.~\ref{figSoftEast} since the active phase (including $s=0$) accommodates the additional transitions (which are not indicated here). Because their thermodynamics is trivial, KCMs do not exhibit the $\varepsilon$-transition of Replica theory, which occurs in configurational space, rather this is limited to systems with non-trivial thermodynamics. For large temperatures, the coexistence line $s_\text{coex}>0$ is again expected to bend away from the temperature axis due to the absence of temperature-independent ``hard'' constraints.

We now review recent work on the role of local structure in dynamical phase transitions and its implications for a possible route to reconcile both theoretical approaches -- dynamic facilitation and the thermodynamic RFOT and Adam-Gibbs.  It is based on the observation that in atomistic models of glass formers kinetic constraints are emerging (coarse-grained) interactions that are necessarily accompanied by a structural signature. This implies (weak) spatial interactions between excitations absent in idealized KCMs. One can therefore probe the role of local structure in the dynamical phase transition exhibited by atomistic models \cite{hedges2009}, and indeed the inactive phase proves to have a higher density of particles in locally favoured structures \cite{speck2012}.

It is also possible to go rather further and to introduce a structural--dynamical order parameter that introduces an explicit structural component to the dynamical phase transition. This  structural--dynamical phase transition is the so-called $\mu$-ensemble, in which distinct phases emerge: poor (active) and rich (inactive) in \emph{time-averaged} populations of structural motifs. In the $\mu$-ensemble, the active phase, like that in the $s$-ensemble, is close to the normal liquid. The inactive phase however is rich in locally favoured structures \cite{speck2012}.

Numerical evidence for the structure-poor and structure-rich dynamical phase coexistence has so far been obtained for three different model glass formers: Kob-Andersen~\cite{speck2012,turci2017prx}, Wahnstr\"om~\cite{turci2018epje} (both are binary mixtures with Lennard-Jones pair potentials), and polydisperse hard spheres~\cite{pinchaipat2017,campo2019}. Generally speaking, similar behaviour is seen in the three models, and we emphasise differences between the models at appropriate points in the discussion.


\vspace{10mm}
\section{Dynamical phase transitions in ensembles of trajectories}
\label{sectionSMu}

\begin{figure}[!htb]
\centering \includegraphics[width=\columnwidth]{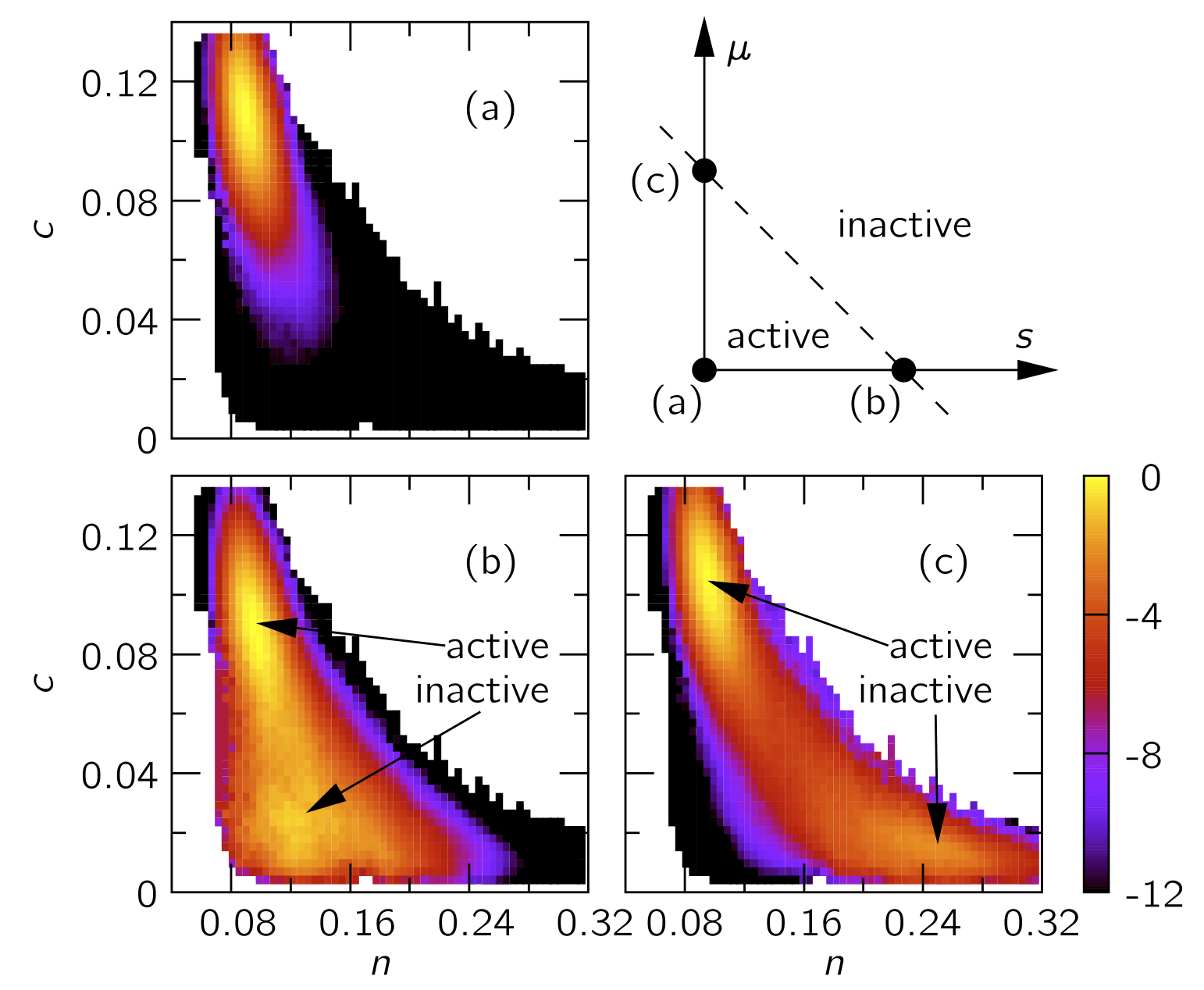} 
\caption{Logarithm of the joint probability for mobility $c$ and time-integrated LFS population $n$ shown for (a) the unbiased ensemble and at coexistence in (b) the dynamical $s$-ensemble and (c) the structural $\mu$-ensemble. The phase diagram in the s-$\mu$ plane is sketched in the upper corner. Reproduced with permission from Ref. \citenum{speck2012}.}
\label{figSMuTogether}
\end{figure}

The purely dynamical $s$-ensemble and structural-dynamical $\mu$-ensemble are constructed in a rather similar manner, and we find it expedient to discuss both together. In the case of the $\mu$-ensemble, the first link that we want to establish is between the dynamical phase transitions of dynamical facilitation and the structural changes observed in the liquid. Is there any relation between the glassy trajectories of the inactive dynamical phase and the emergence of local structural order?

\subsection{The $s$-ensemble}
\label{sectionS}

To answer this question and in the spirit of statistical mechanics, we seek to define order parameters that quantify the corresponding behavior. Specifically, this is
\begin{equation}
  \mathcal C[X] = \sum_{t=1}^{\tobs} N\hat c(t-1,t)
  \label{eq:c}
\end{equation}
measuring the \emph{time-averaged} population $\mathcal C$ of mobile particles \emph{along trajectories} $X$ of length $\tobs$ (which is measured in multiples of the microscopic time required for a single particle to commit to a new position~\cite{keys2013}). Here, $\hat c(t-1,t)$ is the fraction of mobile particles that underwent such a transition to a new (average) position between frames $t-1$ and $t$. We also define the fraction $c=\mathcal C/(N\tobs)$ which takes on values between zero and unity. The crucial feature is that this order parameter is extensive both in space \emph{and} time.

At first glance, as shown in Fig. \ref{figSMu}(a), the probability distribution of $c$ is somewhat unremarkable; they appear as Gaussians as dictated by the central limit theorem, reflecting fluctuations in the supercooled liquid. However, pushing into the tails of the distributions (low values of $\mathcal C$) we soon realize that they decay much slower than what would correspond to a Gaussian [cf. Fig.~\ref{figSMu}(a) for the binary Kob-Andersen mixture]. To elucidate the conceptual behavior, we introduce the external ``field'' $s$,
\begin{equation}
  \mean{\mathcal C}_s = \frac{1}{Z(s)}\mean{\mathcal C e^{-s\mathcal C}},
  \label{eq:rews}
\end{equation}
which promotes rare trajectories with corresponding ``dynamic'' partition function $Z(s)$ ensuring normalization. For obvious reasons this ensemble is dubbed the $s$-ensemble. Equation~\eqref{eq:rews} is a form of importance sampling~\cite{ray2018,jack2019}. The theory behind it connects to the mathematics of large deviations~\cite{touchette2010} and will not be reviewed here. Sampling sufficiently many trajectories in computer simulations at non-vanishing fields is a challenge that requires advanced sampling techniques. Two candidates are transition path sampling~\cite{swendsen1986,bolhuis2002}, and cloning algorithms~\cite{giardina2006,nemoto2016}. Results shown here have been obtained from a combination of transition path sampling with replica exchange, for details see~~\cite{swendsen1986,bolhuis2002}. One consequence of the substantial computational demand is the need to consider small systems composed of a few hundred particles.

\subsection{The $\mu$-ensemble}

Rather than the time-averaged population of \emph{mobile} particles, the $\mu$-ensemble considers time-averaged populations of particles in \emph{locally favoured structures}. The construction of the order parameters and ensemble is entirely analogous to the $s$-ensemble. Following Eq. \ref{eq:c}, the population of structural motifs 
averaged along the trajectory is
\begin{equation}
  \mathcal N[X] = \sum_{t=0}^{\tobs} N\hat n(t),
\end{equation}
where $\hat n(t)$ is the fraction of particles found in the chosen structural motif (typically the locally favoured structure) at $t$.

Similarly to the $s$-ensemble, as shown in Fig. \ref{figSMu}(d), the probability distributions of $n$ is single-peaked, but with a rather clear ``fat tail'' at high $n$, corresponding to much slower decay than would be the case for a Gaussian. As above, we introduce an external ``field'' $\mu$ which promotes rare trajectories,
\begin{equation}
  \mean{\mathcal N}_\mu = \frac{1}{Z(\mu)}\mean{\mathcal N e^{\mu\mathcal N}},
  \label{eq:remu}
\end{equation}
with dynamic partition function $Z(\mu)$. By analogy, we call this ensemble of biased trajectories the $\mu$-ensemble.

\subsection{Structure can drive global dynamics}

Numerical curves of the population of particles in locally favored structures averaged along a trajectory corresponding to Eq.~\eqref{eq:remu} are plotted in Fig.~\ref{figSMu}(c,f) and show that there is a qualitative change as we increase both $s$ and $\mu$ which becomes more abrupt as the trajectory length is increased. The corresponding susceptibilities peak at $s^\ast$ and $\mu^\ast$. These are hallmarks for a phase transition, which here occurs in the space of trajectories.

Figure~\ref{figSMuTogether} shows the joint distribution of $n$ and $c$ for different values of the fields, for which we can discern two basins separated by a barrier. This demonstrates that the transition occurs between a phase that has many mobile particles and a low occupation of LFS (which we identify with the normal supercooled liquid), and a phase with very few mobile particles and a high population of LFS. The later phase shows properties normally associated with a glass. This interpretation is further supported by looking at the reweighted marginal distributions at $s^\ast$ and $\mu^\ast$ [Fig.~\ref{figSMu}(b,e)], which exhibit two peaks. Both phases are also termed active and inactive, respectively. In particular, the $\mu$-ensemble probes the active-inactive transitions explored in the $s$-ensemble: inactive trajectories correlate strongly with trajectories rich in locally favoured structures.

It is worth noting that the change in LFS population $\mean{n}_{s,\mu}$ is rather more marked in the case of the $\mu$-ensemble transition [Fig.~\ref{figSMu}(f)] than is the case for the $s$-ensemble [Fig.~\ref{figSMu}(c)]. While both access the same basin in the mobility--LFS population $(c,n)$ plane (as discussed below, Fig.~\ref{figSMuTogether}), the change in $\mean{n}_{s,\mu}$ between the two ensembles is worthy of some discussion. We make the following observation: the trajectory lengths are up to $\tobs=10\tau_\alpha$ in length, sampled at a temperature of 0.6. For deep supercooling (say $T < 0.4$), the relaxation time is \emph{much} longer than this timescale of $10\tau_\alpha$ at $T=0.6$. Specifically, $\tau_\alpha(T=0.6)\approx 32$, while $\tau_\alpha(T=0.4)\approx 2.91 \times 10^5$ Lennard-Jones time units. So biasing to a low fraction of \emph{mobile} particles is less effective, since the number of mobile particles on the timescale of the trajectories even in the normal liquid is very small for deeply supercooled states. On the other hand, the population of particles in \emph{locally favoured structures} continues to increase significantly even at rather deep supercooling \cite{royall2015,ingebrigtsen2019}, thus biasing on the population of particles in LFS may generate configurations deeper in the energy landscape than biasing on the dynamics.

In particular, we see that in Fig. \ref{figSMuTogether}, the fraction of mobile particles reaches less than 0.01, which, in an $N=216$ particle system, as is the case here, corresponds to just one or two particles being mobile.  Thus the system has, in a sense become almost as slow is it can under the $s$-ensemble biasing -- but, given the short trajectories and small system sizes -- the rate of relaxation is still much higher than would be the case at a much lower temperature where the LFS population would be higher. In other words, under these system sizes (which includes, crucially, trajectory lengths), the system has become almost as slow as it can. In this temperature regime, no such limit pertains to the $\mu$-ensemble, the population of LFS can rise at least as high as $n\approx0.5$, thus applying the biasing field can generate a rather higher LFS population than is the case for the $s$-ensemble.

While an exhaustive range of structures has not (yet) been explored in the $\mu$-ensemble, we remark that the locally favoured structure is identified as that which lasts longer than other candidate local structures that minimise the local energy \cite{malins2013tcc}, as identified in a range of systems \cite{malins2013jcp,malins2013fara,royall2015}. Structures whose symmetry is distinct to the LFS have been investigated and no $\mu$-ensemble type transition was observed \cite{pinchaipat2017}. The possibility of other structures and indeed other metrics such as \emph{order-agnostic} approaches are an intriguing avenue to pursue to investigate other biasing fields for dynamic phase transitions and we return to this point below.

\begin{figure}[t]
\centering \includegraphics{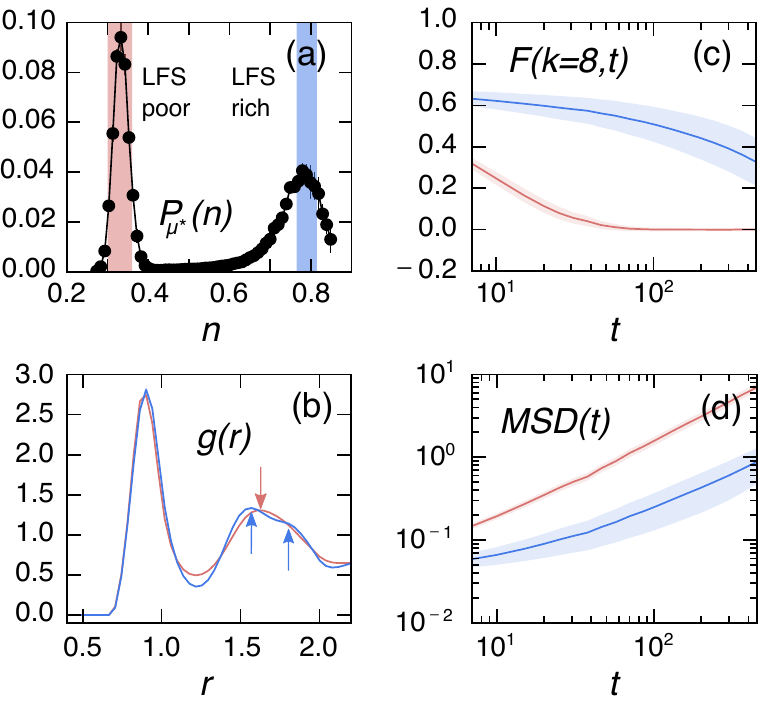}
\caption{Trajectory properties in hard spheres. (a) Distribution function $P_{\mu*}(n)$ of the population of LFS at coexistence. Shaded areas corresponding to the peaks of the distribution indicate the range of $n$ values associated with the normal (active) liquid (LFS-poor, red) and the inactive phase (LFS-rich, blue). (b) Radial distribution functions obtained from trajectories sampled within the LFS-poor (red) and LFS-rich (blue) region of $P(n)$. Arrows indicate the positions of peaks. (c) Intermediate scattering functions and (d) mean-squared displacement for the two phases. Reproduced from [Campo, M. and Speck, T. \emph{J. Chem. Phys.}, \textbf{152} 014501 (2020)], with the permission of AIP Publishing.}
\label{figHardSpheres}
\end{figure}

In Fig.~\ref{figHardSpheres}(a), we plot the reweighted distribution of $n$ for another model system, polydisperse hard spheres~\cite{campo2020}. We now have a closer look at the trajectories populating the first peak (red shaded area, LFS-poor) and the second peak (blue shaded area, LFS-rich). As shown in Fig.~\ref{figHardSpheres}(b), the static structure as measured by the pair distribution function is very similar (note, however, the splitting of the second peak). In contrast, the dynamics as measured by the intermediate scattering function [ISF, Fig.~\ref{figHardSpheres}(c)] and the mean-square displacement [Fig.~\ref{figHardSpheres}(d)] are markedly different, demonstrating that the average dynamics in the LFS-poor phase is ``fast'' while in the LFS-rich phase it is much slower. There is thus again a strong correlation between the overall population of LFS and the global dynamics.

\begin{figure*}
\centering
\includegraphics[width=0.8\textwidth]{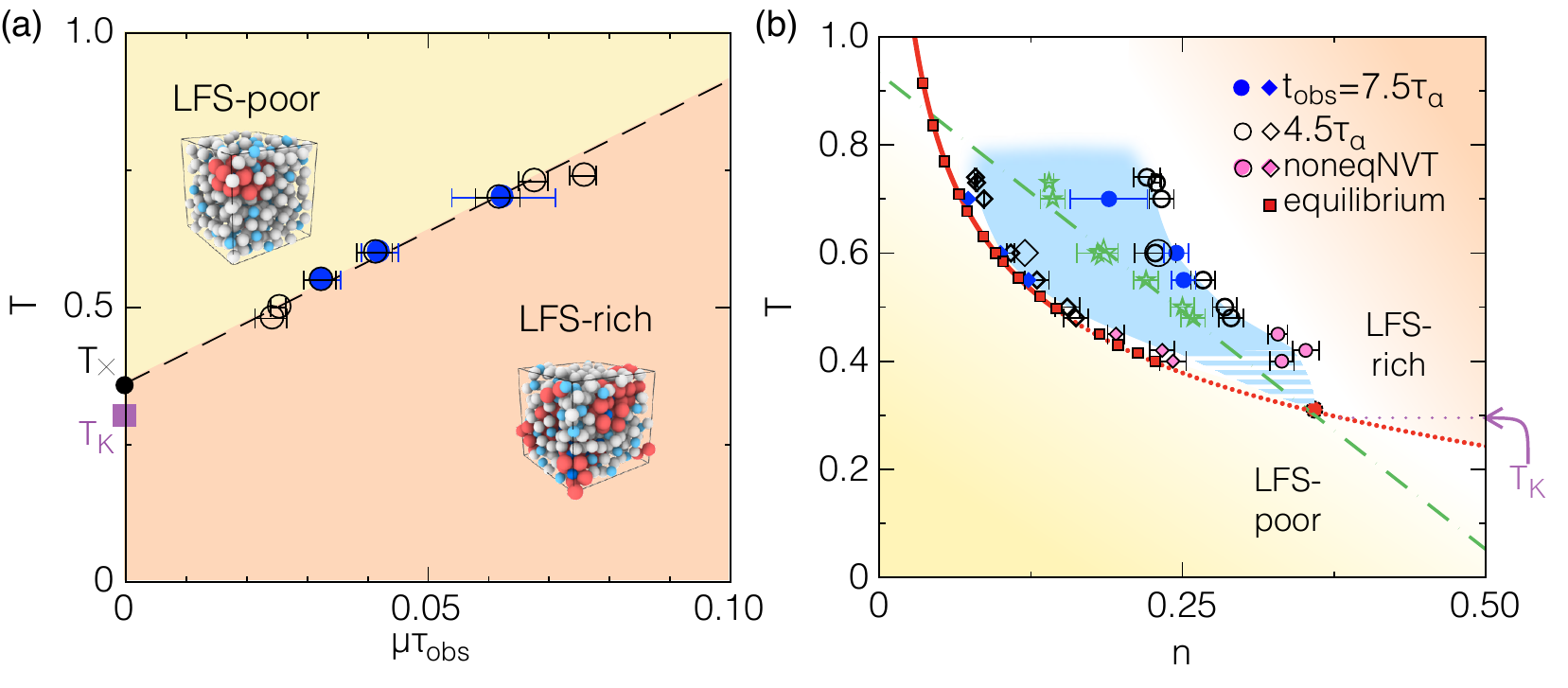}
\caption{Results of trajectory sampling ($\mu$-ensemble) computer simulations of the Kob-Andersen binary Lennard-Jones mixture. (a) Temperature versus $\mu$ phase diagram. Two distinct structural-dynamical phases are found at coexistence at a finite value $\mu^\ast$ of the field $\mu$ when sampling trajectories of different durations $\tobs$ (filled and empty circles): these are poor and rich in structure (LFS poor/rich), represented in the insets (with red and dark blue particles indicating the LFS regions). The scaled value $\mu^{\ast}\tobs$, however, lies on a single line. A linear extrapolation (dashed line) indicates that at a temperature $T_\times$ one would observe the transition from one phase to the other directly in the thermal average of structural quantities ($\mu=0$) without recurring to large deviations, under the form of intermittency. (b) Dynamical coexistence in the temperature versus concentration of LFS per trajectory plane. The coexistence region (determined by several numerical methods, in blue) has a non-trivial temperature dependence and narrows as the temperature is reduced. The equilibrium supercooled liquid approaches the coexistence region gradually and is always located close to the LFS-poor boundary. The extrapolation of the line of susceptibility maxima (green stars) and the equilibrium line meet at a temperature close to $\TK$, suggesting a cross-over between the LFS-poor and the LFS-liquid. More information in Ref.~\citenum{turci2017prx}. Reproduced from Ref.~\citenum{turci2017prx} with additional data from Ref.~~\citenum{ingebrigtsen2019}.}
\label{figMuKA}
\end{figure*}

The phase diagrams that can be constructed from the analysis of probability distributions are presented in Fig.~\ref{figMuKA} for the binary Kob-Andersen mixture, in Fig.~\ref{figMuWahnstrom} for the Wahnstr\"{o}m mixture, and in Fig.~\ref{figHSDynamicalTransition} for a mixture of hard spheres with 10\% polydispersity (spread of particle diameters). All phase diagrams have temperature/inverse density along the $y$-axis and either LFS population $n$ or the conjugate field $\mu$ along the $x$-axis.

Experimental evidence, using colloidal suspensions, has now been found for this structural and dynamical phase transition \cite{pinchaipat2017,abou2018}. More recent studies~\cite{coslovich2016jstat} suggest that the inactive/locally favoured structure-rich phase obtained in the space of trajectories also correlates strongly with particularly low (potential) energy states, and that when decreasing the temperature, the inactive, LFS-rich phases tend to dominate the statistics, with the coexistence values of $s^{\ast}$ and $\mu^{\ast}$ approaching zero as the temperature is decreased. In this sense, guiding trajectory sampling with the usage of time-integrated observable can be an efficient way to identify low energy states, more present in the arrested glassy phases.

\begin{figure*}[t]
\centering \includegraphics[width=0.6\textwidth]{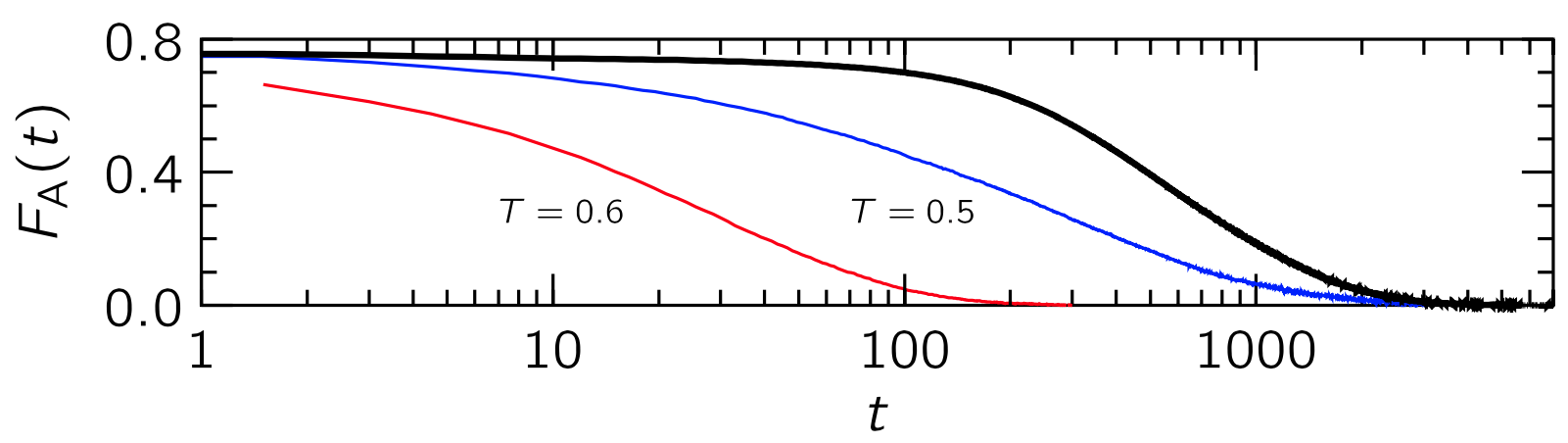}
\caption{``Melting'' glassy configurations generated with the $\mu$--ensemble. The intermediate scattering function is shown for an inactive LFS-rich configuration, run at the sampling temperature $T=0.6$ (black line). Although it eventually decays, indicating particle movement, the ISF $(F(k,t))$ remains close to unity for an extended period, indicating that the particles do not move. Here the wavevector for the ISF is close to the main peak of the structure factor. This relaxation is much slower than the equilibrated supercooled liquid at $T=0.6$ (red line) and even the liquid at $T=0.5$, \emph{below} the sampling temperature is much quicker to relax than the  $\mu$--ensemble configuration. Reproduced with permission from Ref.~\citenum{speck2012}.}
\label{figMuISF}
\end{figure*}

Before discussing the trajectory reweighting \cite{turci2017prx}, which enables one to generate configurations representative of very deeply supercooled states, we directly address the title of this section, \emph{structure can drive global dynamics}. Before proceeding, we note that the intriguing dynamical behaviour of the isoconfigurational ensemble of Harrowell and coworkers \cite{widmercooper2006}, where different regions of the system exhibit consistently different mobility, when run from the same configuration with randomised velocities. This provides strong evidence in support of the idea that certain configurations lead to slower dynamical behaviour than others. In other words, dynamic heterogeneity is encoded in the structure in the atomistic systems considered.

We now discuss evidence for the effect of the configurations generated by the trajectory sampling of the $s$-- and $\mu$--ensembles 
on the dynamics of the system. In the case of KCMs, Keys \emph{et al.} showed that the $s$--ensemble produced inactive configurations with properties representative of states at far deeper supercooling that the temperatures at which they had been sampled \cite{keys2015}. For atomistic systems, using the $\mu$-ensemble, Speck \emph{et al.} showed that inactive configurations were effectively solids which ``melted'' after a certain incubation time when run with unbiased dynamics at the sampled temperature ($T=0.6$) \cite{speck2012}. As shown in Fig. \ref{figMuISF}, even the unbiased liquid at a lower temperature ($T=0.5$) had a more quickly decaying intermediate scattering function than a configuration biased in the $\mu$-ensemble. Qualitatively similar results were obtained for the $s$--ensemble by Jack \emph{et al.} \cite{jack2011}. We conclude then, in addition to the dynamical phase transition and its place in the facilitation picture that, regardless of any theoretical standpoint, the biasing $s$-- and $\mu$-ensembles generate configurations that are more ``solid--like'' than those run with conventional dynamics.

\section{Reweighting for deeply supercooled configurations}
\label{sectionReweighting}

Guiding trajectory sampling has been recently tested in the case of a canonical atomistic glass former (the Kob-Andersen binary mixture) \cite{turci2017prx}, and it has been shown that other systems exhibit similar behaviour, such as hard spheres \cite{pinchaipat2017} and the Wahnstr\"{o}m binary Lennard-Jones model \cite{turci2018epje}. In particular, it has been shown that the large deviations of time-integrated structural observables give access to configurations that sample the tails of the probability distribution of inherent state energies, which appear consistent with those sampled at lower temperatures than accessed directly in the simulations. \emph{Under the assumption that these configurations are indeed representative of the system at low temperature}, we now consider \emph{reweighting} to recover the thermodynamical properties of the system (such as the configurational entropy) down to very low temperatures, without the need of sampling the dynamics at low temperatures directly. Before proceeding, we emphasise that this method enables us to access configurations representative of states very much more deeply supercooled than the temperatures at which we sample. For example, in the case of the Kob-Anderson model, whose mode-coupling crossover is $T_\text{MCT}\simeq0.435$ and the temperature obtained by fitting the Vogel-Fulcher-Tamman equation $T_\text{VFT}\approx\TK\approx0.30\pm0.02$ \cite{sciortino1999,turci2017prx,ingebrigtsen2019,coluzzi2000}, the lowest sampling temperature is 0.48, yet reweighting provides access to configurations representative of the system at temperatures of $T\approx0.35$ or even less~\cite{turci2017prx}.

Our measure of configurational entropy is via the number of amorphous inherent states $\Omega(\phi) \propto e^{N \sigma(\phi)} \delta \phi$ in a range of inherent state energy per particle $\phi - \delta \phi/2 \leq \phi < \phi + \delta \phi/2$. Here $\sigma(\phi)$ is the enumeration function which is quadratic in $\phi$. Thus sampling of configurations with very low inherent state energies via the $\mu$-ensemble gives a measure of the density of states as a function of the inherent state energy. In the thermodynamic limit, the extensive configurational entropy becomes $\ln(\Omega)\approx NS_\mathrm{conf}$ with $S_\mathrm{conf}=S_\infty+\sigma(\phi)$ above the thermodynamic (Kauzmann) transition and $S_\mathrm{conf}=0$ below. We obtain a configurational temperature from $d\sigma/d\phi=1/T_\text{conf}$. Further details, along with methods to reweight to the case that the effective biasing field is removed and thus obtain configurations representative of the experimental $(\mu=0)$ case are given in ref.~\citenum{turci2017prx}.

Additionally, this approach shows that the dynamical phase transition between trajectories poor/rich in local structure sampled in the trajectory ensemble corresponds to a transition between two distinct metastable amorphous states at high/low inherent state energies respectively: one corresponds to the supercooled liquid sampled in conventional dynamics; the other to a secondary amorphous state, with low energy, low configurational entropy, rich in structure and very slow dynamics, see Fig.~\ref{figMuKA}. This second amorphous state is more metastable than the conventional supercooled liquid, however, the difference in stability (as measured by the value $\mu^\ast$ of the conjugated field $\mu$ at coexistence between the two phases) is a function of the temperature and \textit{decreases} as the temperature is reduced.

\begin{figure}
\centering
\includegraphics[width=80mm]{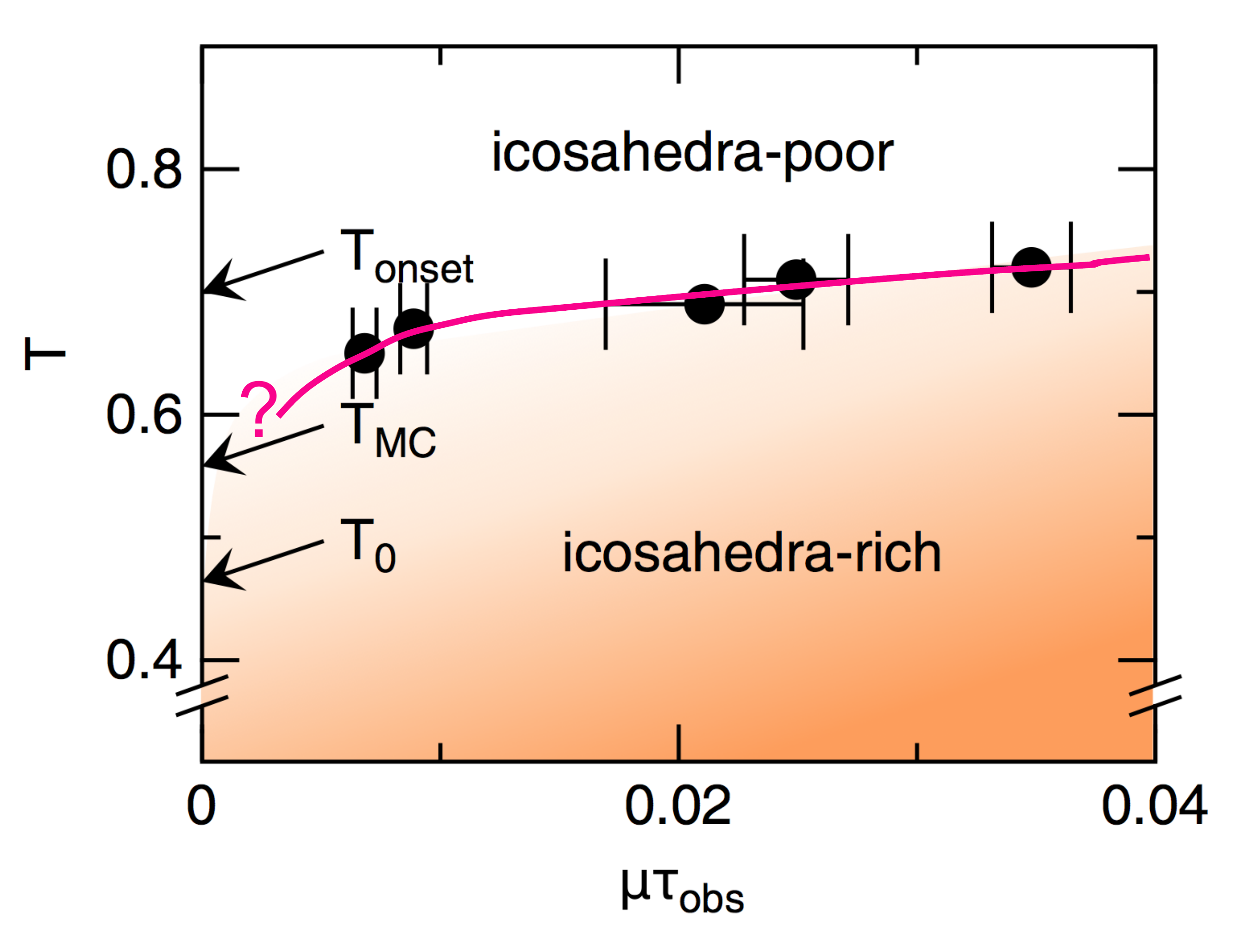}
\caption{Results of trajectory sampling ($\mu$-ensemble) computer simulations of the Wahnstr\"{o}m binary Lennard-Jones mixture. In this system, the LFS is the icosahedron \cite{coslovich2007,malins2013jcp}. Temperature versus $\mu$ phase diagram. A structural-dynamical phase coexistence is found at  at a finite value $\mu^\ast$ of the field $\mu$ when sampling trajectories of durations $\tobs$. These are poor and rich in structure (LFS poor/rich). Pink line is to guide the eye. Reproduced with kind permission of The European Physical Journal E (EPJE) Ref.~\citenum{turci2018epje}.}
\label{figMuWahnstrom}
\end{figure}

Extrapolations are consistent with the scenario that $\mu^\ast(T)\rightarrow0$ at a finite crossover temperature $T_\times\gtrsim \TK$. However, we emphasise that between the lowest temperature at which we sampled, $T=0.48$, and the temperatures to which the system is reweighted, the relaxation time increases to an enormous extent. Moreover,  in the case of the other models which have been investigated, in particular the Wahnstr\"{o}m binary Lennard-Jones model \cite{turci2018epje}, the extrapolation does not lead to $T_\times\approx\TK$, as $\mu^\ast(T)\rightarrow0$, and that indeed  $\mu^\ast(T)$ does not seem to follow a straight line (Fig. \ref{figMuWahnstrom}). The same holds for hard spheres as shown in Fig. \ref{figHardSpheres} \cite{campo2020}. The observation is that while the topology of the phase diagram is preserved, the actual degree of metastability of the supercooled liquids, as quantified by the coexistence value of the dynamical chemical potential as a function of temperature, actually depends on the details of the system. Yet, we find that the emerging, long-lived LFS in the structure-rich phase has a direct physical meaning: configurations extracted from the structure-rich phase display a more rigid response, related to the enhanced stability of the locally favoured structures~\cite{turci2018epje}.

\begin{figure}
\centering
\includegraphics{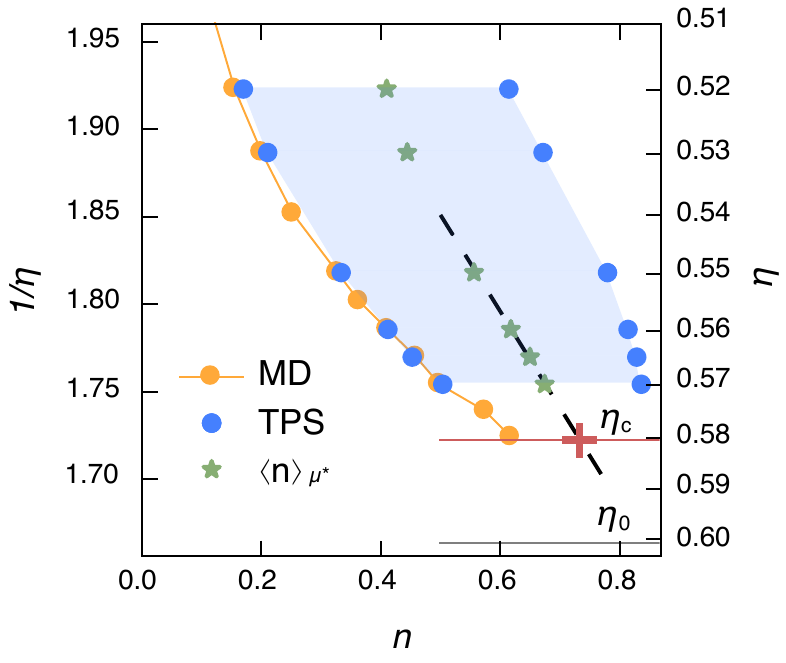}
\caption{The $\mu$-ensemble coexistence region for hard spheres in the ($1/\eta$,$n$) plane, where $\eta$ is the volume fraction, represented as the blue shaded area between the active phase ($n^\mathrm{act}$) and inactive phase ($n^\mathrm{in}$) peaks from the dynamical phase transition. Results are shown for biased transition path sampling (TPS) simulations (blue symbols) together with the population in the liquid phase obtained from straightforward simulations [orange symbols, also shown in Fig. 1(b)]. The dashed line is a linear fit of the ensemble-averaged population at coexistence (green, star symbols), $\eta \geq 0.55$. The gray horizontal line indicates the location of $\eta_0$ where the VFT expression for the relaxation time $\tau_\alpha$ diverges. The red horizontal line indicates the location of $\eta_c$ where the barrier between phases is extrapolated to become independent of $\tobs$. Modified from [Campo, M. and Speck, T. \emph{J. Chem. Phys.}, \textbf{152} 014501 (2020)], with the permission of AIP Publishing.}
\label{figHSDynamicalTransition}
\end{figure}

\textit{Is the reweighting necessary?} 
Before moving to put the pieces together and constructing our standpoint we pause to consider the importance of the reweighting methods we have discussed \cite{turci2017prx,turci2018epje}. While these enable us to access configurations inaccessible to brute force simulations, due to their low temperature, in fact it is possible to identify the dynamical phase behaviour we consider \emph{without} reweighting \cite{campo2020}. The results are shown in Fig. \ref{figHSDynamicalTransition}. It is clear in this figure that the LFS-poor (normal liquid) and LFS-rich (inactive) phases approach one another in much the same way as is the case of the reweighted data used in the case of the Lennard-Jones models above \cite{turci2017prx,turci2018epje}. We thus conclude that, important though the reweighting is to access deeply supercooled configurations, it is not necessary to demonstrate the topology of the dynamical phase transition.

\section{Putting it all together}
\label{sectionAllTogether}

\begin{figure*}
\centering
\includegraphics[width=\textwidth]{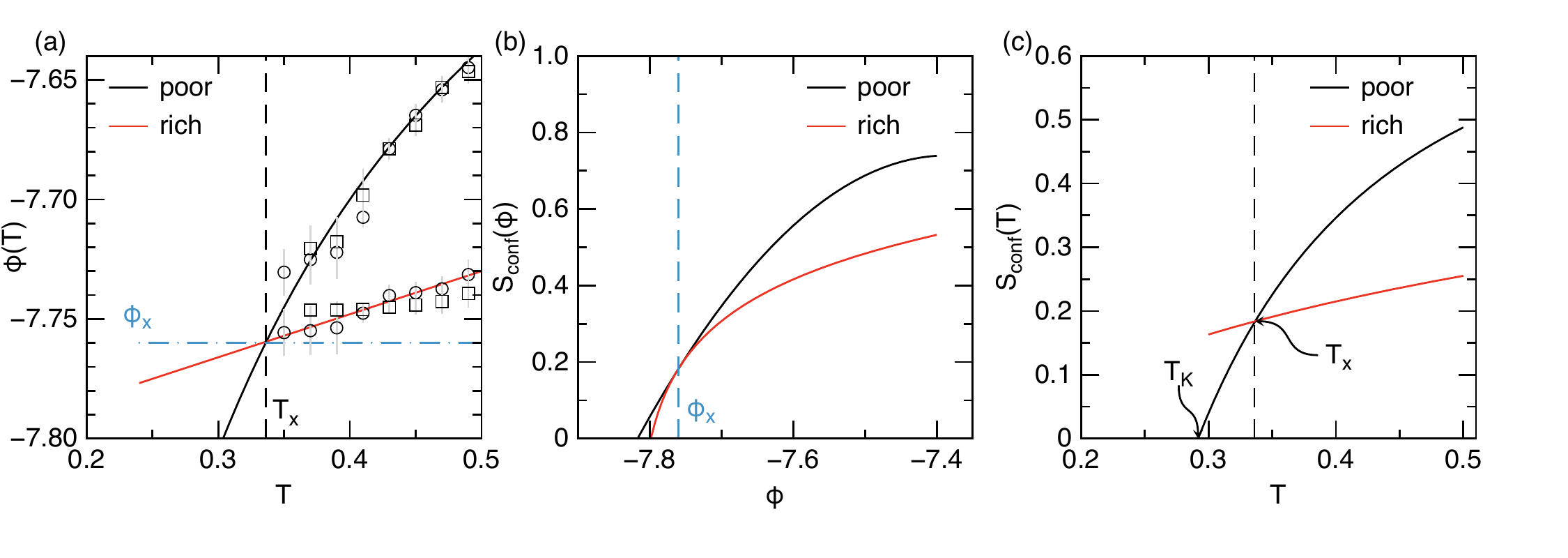}
\caption{(Colour online) Schematic picture of a tentative continuous transition. Shown are the trajectory sampling simulations for the binary Kob-Andersen mixture of Ref.~\citenum{turci2017prx} reweighted to the unbiased dynamics $\mu=0$. (a) Inherent state energy per particle as a function of temperature, (b) entropy as a function of inherent state energy, and (c) entropy as a function of temperature. In (a) circles and squares correspond to the LFS poor/rich inherent state energies at coexistence obtained for two sampling temperatures $\Ts=0.50, 0.48$.}
\label{figEntropyISESchemes}
\end{figure*}

We have shown that the dynamical phase transition in the $\mu$-ensemble, albeit of a (time-integrated) structural quantity, allows access to states very deep in the energy landscape and in the case of the KA model seems to have a lower critical point close to the putative Kauzmann transition of Adam-Gibbs and RFOT theory~\cite{turci2017prx}, as indicated in Figs. \ref{figCavagnaWDIAM} and \ref{figEntropyISESchemes}(c). How can it be that a method based on \emph{dynamical} phase transitions, starting from such a wildly different standpoint, can apparently start to relate to the \emph{thermodynamic} Kauzmann-type scenario?

Some insight may be gleaned from examining certain physical quantities in the two phases -- the structure-rich inactive phase and the structure-poor, active phase, which lies close to the normal unbiased, supercooled liquid (in the sense that $n^\mathrm{act}\approx\mean{n}$). We begin with the inherent state energy $\phi$, in Fig.~\ref{figEntropyISESchemes}. It is clear that, at relatively high temperature (e.g. $T=0.6$ in the Kob-Andersen model), the typical inherent state energy of the structure-rich inactive phase is very low (relative to the structure-poor equilibrium liquid). Upon dropping the temperature, in the structure-rich inactive phase, the inherent state energy decreases more gradually upon reducing the temperature, while that of the structure-poor active phase falls markedly.

In the case of the Kob-Andersen mixture, the simulations in trajectory space~\cite{turci2017prx} allow us to be more quantitative. The average inherent state energy of the structure-poor liquid is found to be well modelled by
\begin{equation}
  \phi^{\rm act}(T)=\phi_{\infty}-\frac{J^2}{2T}
  \label{eq:phipoor}
\end{equation}
in the regime where inherent states are well defined (i.e. for temperatures well below the onset of slow dynamics), with $J$ being a fitted characteristic energy scale. In the same regime, the average inherent energy of the structure-rich states follows approximately
\begin{equation}
  \phi^{\rm in}(T)\approx \gamma T+\phi_0,
  \label{eq:phirich}
\end{equation}
with fitting parameters $\gamma$ and $\phi_0$. These specific forms suggest the existence of a crossover temperature $T_\times$ at which the structure-poor and the structure-rich state become indistinguishable, which for the particular values of the fitting parameters results to be $T_\times\approx0.33$. This is a suggestive insight of the trajectory sampling approach in the context of atomistic glassformers: the structural-dynamical transition appears to terminate at a very low temperature in a critical point close to, but above, the estimates of the temperature at which the relaxation times diverge (the Vogel-Fulcher-Tamman fit used here suggests a divergence at $T_\text{VFT}\approx0.325$), though as noted above, a range of values have been obtained for such a substantial extrapolation and we take $T_\text{VFT} \approx \TK \approx 0.30 $ \cite{sciortino1999,,turci2017prx,ingebrigtsen2019,coluzzi2000}.

As the inherent state energies of structure-rich and poor states converge, we can follow the decrease of configurational entropies of the two disordered states with decreasing temperature. In particular, configurational entropy $S_\mathrm{conf}$, inherent state energy $\phi$, and as noted in section \ref{sectionReweighting}, (configurational) temperature are related by $dS_\mathrm{conf}/d\phi=1/T$. Thus the two equations~(\ref{eq:phipoor},\ref{eq:phirich}) imply $ dS_\mathrm{conf}^{\rm act}/d\phi^{\rm act}=2(\phi_{\infty} -\phi^{\rm act} )/J^2$ and $dS_\mathrm{conf}^{\rm in}/d\phi^{\rm in}=\gamma/(\phi^{\rm in}-\phi_0)$ respectively.

After integration
\begin{equation}
  S_\mathrm{conf}^{\rm act}(\phi) = S_\mathrm{conf}^\infty-\frac{(\phi-\phi_{\infty})^2}{J^2},
\end{equation}
and
\begin{equation}
  S_\mathrm{conf}^{\rm in}(\phi) =\gamma \ln \frac{\phi-\phi_0}{G},
\end{equation}
where $G$ is an integration constant. Imposing that at $T_\times$ the two states also have the same configurational entropy fixes the value of $G$. Figure~\ref{figEntropyISESchemes}(b) illustrates the two branches of configurational entropies for the two states. The same information can be cast as a function of temperature, see Fig.~\ref{figEntropyISESchemes}(c), where the decrease in the entropy difference between the structure-poor and structure-rich states as a function of the temperature is explicit. We remark that Fig.~\ref{figEntropyISESchemes}(c) is reminiscent of the original Kauzmann plot of (configurational) entropy shown in Fig. \ref{figCavagnaWDIAM}, where the structure-rich, low entropy state plays now the role originally assigned to the crystal, albeit with a slightly higher configurational entropy.

\emph{What does it all mean? --- }
The emerging picture is that at a low enough temperature, the equilibrium, structure-poor supercooled liquid becomes indistinguishable from a very low entropy, very low energy structure-rich metastable state. Remarkably, this picture contains elements of both the Adam-Gibbs/RFOT scenario (i.e. a steadily decreasing configurational entropy to a disordered low entropy state at very low temperatures) and the dynamical facilitation scenario indicated in Fig. \ref{figCavagnaWDIAM} (i.e. the structure-rich state is at the same time an inactive state \cite{speck2012jcp,speck2012}).

An important question revolves around the location and nature of the crossing point $T_\times$ of the inherent state energy and configurational entropy. This is ultimately related to the fate of the structural-dynamical phase coexistence in trajectory space, and in particular the location of any lower critical point of the dynamical phase transition. Several possibilities arise (see more detailed discussions in Refs. \citenum{turci2017prx} and \citenum{royall2018jpcm}). Simulation results of three particulate systems (the Kob-Andersen binary mixture, the Wahnstr\"om mixture, and polydisperse hard-spheres) indicate that the coexistence terminates at temperatures between the Kauzmann temperature $\TK$ and the mode-coupling crossover $T_\text{MCT}$. Whether the coexistence terminates at nonzero dynamical chemical potential $\mu_\text{c}$ is still an open question, whose resolution will depend on more accurate, low temperature measures and finite size studies. We recall that the $\mu$-ensemble is explicitly a \emph{structural-dynamical phase transition, so any divergence at a lower critical point is expected in space as well as time. We believe that this character should also be exhibited by the $s$-ensemble, given the change in local structure across the $s$-ensemble (Figs. \ref{figSMu} and  \ref{figSMuTogether}). Therefore we expect that any lower critical end-point would be accompanied by a diverging static correlation length influencing the unbiased liquid even for $\mu_\text{c}>0$ and providing a mechanism for increasing static correlations.}

As to the exact relationship between the dynamical phase transition and any thermodynamic glass transition it is clear that the structure-rich inactive phase is a state with exceedingly low configurational entropy, and so in some sense lies at least close to any ``ideal glass'' state with vanishing configurational entropy.

How does this square with the kinetically constrained models?  Interestingly, by adding ``softness'', i.e. by softening the constraints of the East model (a KCM), Elmatad and Jack were able to show a profound difference in its dynamical phase diagram (Fig. \ref{figSoftEast}) \cite{elmatad2013}. As noted above, the unmodified East model has its dynamical phase transition at $s=0$, but the softness led to a lower critical point reminiscent of that in the Kob-Andersen model in Fig. \ref{figMuKA}. The dynamical phase transition moreover shifted to a positive value of the dynamical field $s$. Related results were obtained by Turner \emph{et al.} \cite{turner2015} with placquette models, which might also be thought of as ``KCMs with thermodynamics''. Here the same model showed the $s$-ensemble type transition, and the $\varepsilon$-coupling of the Replica theory. Moreover, as already noted above, spin glasses with non-trivial thermodynamics \cite{jack2010} can also exhibit a dynamical transition.

The picture that emerges is that in systems with nontrivial thermodynamics, be they atomistic or colloidal glassformers \cite{turci2017prx,turci2018epje,campo2020}, or spin-glasses \cite{jack2010}, the dynamical phase transition has a lower temperature end-point (critical point) at finite temperature. Such behaviour is supported in KCMs through softening constraints \cite{elmatad2013}.

We summarise our standpoint as follows.
\begin{itemize}
\item{The $\mu$-ensemble dynamical phase transition has two branches (active and inactive) which approach one another at low temperature.}
\item{Assuming a full convergence, there should be a structural-dynamical critical point with diverging length- and time-scales.}
\item{The low configurational entropy of the inactive phase is reminiscent of the crystal in the Kauzmann plot.} 
\item{Since the inactive phase is amorphous, $S_\mathrm{conf}^\mathrm{in}>S_\mathrm{conf}^\mathrm{xtal}$ so any convergence with the normal liquid should occur for $T>\TK$ (Fig. \ref{figCavagnaWDIAM}). This holds if any lower critical point occurs at $\mu=0$, which itself is not guaranteed.}
\item{A lower lower critical point which occurs at $\mu>0$ would correspond to an avoided transition, which may nevertheless lead to large static lengthscales in the spirit of Geometric Frustration \cite{tarjus2005}.}
\end{itemize}

\section{Challenges and Outlook}
\label{sectionChallengesOutlook}

An obvious numerical challenge is to increase the system sizes that can be addressed with the current method. Sampling fluctuations becomes exponentially more expansive as the number $N$ of particles and the length $\tobs$ of trajectories is increased. In itself this is not a fundamental problem as finite-scaling is a valuable tool in computational statistical mechanics that allows to systematically extrapolate the thermodynamic limit behavior. Still, for reliable finite-size scaling one would like to cover at least on order of magnitude in both $N$ and $\tobs$, which is still out of reach at the moment. One step in this direction has been taken recently for polydisperse hard spheres~\cite{campo2020}, where the trajectory length has been varied. Complementary numerical methods such as population dynamics~\cite{nemoto16} might be helpful here.

Recently, it has been shown that popular model glass formers like the binary Kob-Andersen mixture discussed here crystallize in very large simulations~\cite{toxvaerd2009,ingebrigtsen2019}. The mechanism is through spontaneous composition fluctuations that yield domains of one species large enough to overcome the nucleation barrier to crystallization. Such compositional changes are not accessible in the small systems we studied. It is possible that the dynamic phase diagrams presented in Fig.~\ref{figEntropyISESchemes} are \emph{metastable} with respect to crystallization. This is not a fundamental limitation since virtually any physical supercooled material is metastable with respect to crystallization, and our aim is to gain insight into the vitrification mechanism.

A related issue is that the buildup of structural correlations in the structure-rich inactive state has repercussions on the orientational correlations as well. Since only relatively small systems of some hundreds of particles can--with the present numerical methods--efficiently sample the structural-dynamical transition at low temperatures, the emerging orientational correlations involve the entirety of the sampled regions of space once the structure-rich state is accessed. It is interesting to note, however, that finite-size and compositional constraints prevent structure-rich configurations from forming complex equilibrium crystalline structures such as the Laves phases of polydisperse hard spheres \cite{bonnimeni2019} or crystallising like binary glass formers \cite{ingebrigtsen2019}. Nevertheless, the proximity of these structural transitions can be expected to shape the dynamic phase diagram.

In the context of connecting to the Kauzmann paradox of the converging configurational entropy of a supercooled liquid and its crystal at low temperature, it would be interesting to explore the methods outlined in a system with a well-defined crystal. We expect that the active-inactive dynamical phase transition would be bounded by the liquid-crystal lines in the temperature-configurational entropy plane as sketched in Fig. \ref{figCavagnaWDIAM}. While this may seem challenging with some of the models reviewed here as they have no known crystals of the same stochiometry of the system, the tantalising promise of a model system with a well-defined local structure and crystal whose configurational entropy could be evaluated would be a most interesting prospect. One possibility is metallic glassformers, represented through the embedded atom model. These are reasonably resistant to crystallisation \cite{ingebrigtsen2019}, and the crystal phase diagram has been determined \cite{tang2012}.

Clarifying the relationship between local structure, local configurational entropy and mobility excitations appears as a key task for a more complete theory of dynamic arrest. This will include developing a systematic framework for coarse-graining the model-specific aspects and predicting physically relevant quantities, such as the activation energies advocated by the dynamical facilitation or the size and shape of the cooperatively rearranging regions of the RFOT/Adam-Gibbs scenario. Ensembles analogous to the $\mu$-ensemble can be devised for other quantities which have been defined to quantify glassy systems, including soft spots \cite{schoenholz2014,zylberg2017}, aggregated softness fields \cite{schoenholz2016}, two-body excess and patch entropy \cite{sausset2011,hallett2018,ingebrigtsen2018}, local bonding and packing \cite{tong2018} and community inference \cite{paret2020} and may help to elucidate the relationships between these different descriptors of glassy heterogeneities and their relationship with the dynamics.


\section{Conclusions}
\label{sectionConclusions}

The notion of metastability implies local equilibrium on finite timescales \cite{sewell1980}. Supercooled liquids are metastable disordered states, with complex energy landscapes whose topology is believed to influence the emergent relaxation patterns \cite{debenedetti2001}. Here we have revisited recent results that connect purely dynamical and thermodynamical descriptions of glassy behaviour within a particular framework designed to deal with metastable states, i.e. the theory of large deviations of structural and dynamical observables. The key outcome is that over suitably long observation timescales the dynamics of supercooled liquids  explores trajectories that can be characterised either by high mobility and modest structural order or low mobility and enhanced local structural features. Interestingly, a first-order transition in trajectory space can be associated with this behaviour, and it is common to different models of structural glasses, i.e. additive and non-additive Lennard-Jones mixtures and purely repulsive size-dispersed hard spheres.

Most importantly, the transition is strongly affected by decreasing the temperature: at lower and lower temperatures, the inactive, structure-rich trajectories are less and less distinguishable from the active, structure-poor ones. A characterisation of the energy landscape explored by the two dynamical phases shows that while the active trajectories sample relatively high energy and entropy regions, the inactive ones explore a narrow region of low energy and low entropy. This observation of the merging of the two dynamical phases at low temperature enables us to suggest that it may be possible to bring together the dynamical phase transition of dynamic facilitation with the mosaic of low entropy regions of the Adam-Gibbs/RFOT scenario. The population of locally favoured structures per trajectory is then used as a reaction coordinate to explore metastability, as it couples at the same time with inherent state energies and particle mobility.

Considering the ``Kauzmann plot'' (Fig. \ref{figCavagnaWDIAM}), we see that the inactive phase plays a role similar to that of the crystal, with a small configurational entropy which slowly reduces as a function of temperature. Small as it is, the configurational entropy of the inactive phase is somewhat higher than that of the crystal, so merging of the two dynamical phases is expected at a temperature higher than $\TK$. A lower critical point where the phases merge would have structural and dynamical characteristics, such as diverging time- and length-scales, which are also anticipated in thermodynamic theories of the glass transition.

\acknowledgments

The authors would like to acknowledge Matteo Campo, Alex Malins, Rattachai Pinchaipat, and Stephen Williams for their contributions to the various stages that have cumulated in this Perspective. Without necessarily implying their agreement with everything that is written here, Rob ``P'' Jack, Juan P. Garrahan, and David Chandler are warmly thanked for very many illuminating conversations. We are grateful to Daniele Coslovich for helpful comments on the manuscript. CPR acknowledges the Royal Society. CPR and FT gratefully acknowledge the European Research Council (ERC consolidator grant NANOPRS, project number 617266) for financial support.

\section*{Data Availability Statement}

The data that support the findings of this study are available from the corresponding author upon reasonable request.

\section*{References}


\begin{thebibliography}{100}
\bibitem{}
The devil is in the details: pentagonal bipyramids and dynamic arrest james e
  hallett1,2, francesco turci1 and c patrick royall1,2,3.
\end{thebibliography}

\end{document}